\documentstyle[sprocl]{article}

\input{psfig}

\def\NPB#1#2#3{{\it Nucl.\ Phys.}\/ {\bf B#1} (19#2) #3}

\def\PRD#1#2#3{{\it Phys.\ Rev.}\/ {\bf D#1} (19#2) #3}
\def\PRL#1#2#3{{\it Phys.\ Rev.\ Lett.}\/ {\bf #1} (19#2) #3}

\def\be{\begin{equation}}
\def\ee{\end{equation}}
\def\bea{\begin{eqnarray}}
\def\eea{\end{eqnarray}}

\newcommand{ \ltap }{\stackrel{\lower.65ex\hbox{$<$}}
                     {\lower.65ex\hbox{$\sim$}}}
\newcommand{ \gtap }{\stackrel{\lower.65ex\hbox{$>$}}
                     {\lower.65ex\hbox{$\sim$}}}
\newcommand{ \gsim }{\mathrel{\gtap}}
\newcommand{ \lsim }{\mathrel{\ltap}}

\begin{document}

\title{SUPERSYMMETRIC ELECTROWEAK BARYOGENESIS AND CP VIOLATION}

\author{MICHAL BRHLIK}

\address{Randall Physics Lab, University of Michigan\\
Ann Arbor, MI 48109-1120, USA\\E-mail: mbrhlik@umich.edu}


\maketitle\abstracts{I review the mechanism of electroweak baryogenesis
within the framework of a string motivated supersymmetric extension of
the Standard Model with large flavor-independent CP-violating phases.
Possible implications for the Higgs sector are considered in correlation
with the properties of the right-handed stop. I also comment on the
compatibility of supersymmetric electroweak baryogenesis with various 
CP-violating observables.  
}

One of the problems of the Standard Model is the long standing issue of the 
origin of CP violation. The only source of CP violation in the  Standard Model
comes from the phase $\delta$ in the CKM quark mixing matrix. Once this
phase is determined from experiment it is, at least in principle,
possible to make predictions for all CP-violating observables based on
this one measurement. The Standard Model can then satisfy the necessary
criteria -- baryon number violation, C and CP violation and 
non-equilibrium conditions \cite{sakh} -- allowing for a non-zero baryon asymmetry
to be generated in the universe at the electroweak transition. The 
baryon number violating interactions are introduced by non-perturbative
weak sphaleron decays and the electroweak phase transition provides 
out-of-equilibrium conditions needed to preserve the generated asymmetry.

At high temperatures, the vaccum expectation value of the 
Higgs field is zero but as the universe cools down over time, a second
minimum appears in the potential at $v \neq 0$. As the temperature decreases, 
the probability of making a transition from symmetric to broken phase
grows  and when the transition occurs at a certain point in space, 
a bubble of broken phase forms and expands. The baryon asymmetry is
generated as the wall of the expanding bubble
passes through points in space. Particles in the unbroken phase close 
to the wall interact with the changing Higgs field profile in the
wall and the presence of CP-violating couplings produces source terms
for participating particles. Different chiralities couple with different
strength when CP is violated and a difference occurs in the
reflection and transmission probabilities for the two different chiralities.
Due to rapid gauge, Yukawa and 
strong sphaleron interactions, the CP-violating source terms are 
translated into a net left handed weak doublet quark density which is
finally converted into a baryon asymmetry by weak sphaleron decays. The
non-perturbative asymmetry then diffuses through the bubble wall into
the broken phase where the weak sphaleron interactions are exponentially 
suppressed. Subsequent washout of the baryon asymmetry can therefore be
kept under control provided the first order phase transition is strong 
enough \cite{nel}.  

Unfortunately, this scenario is not viable in the Standard Model since
a small value of the vacuum expectation value at the critical
temperature $v(T_c)/T_c<1$ allows for the baryon asymmetry to be
washed out unless the Higgs mass is below 
$60\,\rm GeV$ which is in direct contradiction with experimental limits.
The situation can be improved in the Minimal Supersymmetric Standard
Model (MSSM) as the light right handed top squark contribution to the
temperature dependent effective potential can push the value of
$v(T_c)/T_c$ up to acceptable values $\geq 1$ even for light Higgs
masses allowed  by experimental searches.

The CP-violating interactions in the MSSM arise in the complex phases of
the soft supersymmetry breaking terms in the Lagrangian and in the phase
of $\mu$. These phases have to be small, typically $\lsim 10^{-2}$, if
they are considered individually with sparticle masses $\cal O$(TeV),
otherwise they induce contributions to the electric dipole moments
(EDMs) of the neutron and electron exceeding experimental limits. 
Recently it has been shown, however, that these stringent limits can be
avoided in the Type I string models with non-universal gaugino masses 
as a result of a particular embedding of the Standard Model gauge group
into two D-brane sectors \cite{bgk,bekl}. The relations among soft
breaking parameters
ensure cancellations of individual contributions to the EDMs and viable
large phase solutions can be obtained over a wide range of parameter
space as illustrated \cite{bekl} in Fig.1. Moreover, the two gaugino
mass phases $\varphi_1=\varphi_3$ (in our parametrization $\varphi_2=0)$, 
the $\mu$ parameter phase $\varphi_{\mu}$ and the 
overall trilinear parmeter phase $\varphi_A$ are {\it flavor-independent}
at the string scale making it possible to separate the physics of CP
violation from flavor physics.

\begin{figure}[t]
\centerline{\psfig{figure=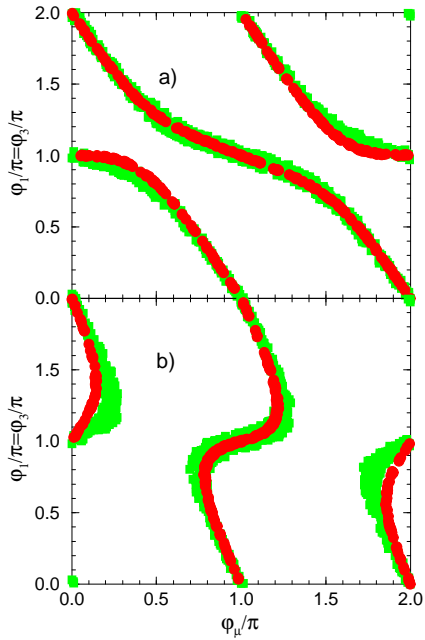,height=3.5in}}
\caption{ 
Electron (red (black) circles) and neutron (green (grey)
blocks) EDM allowed regions for the Type I
orientifold models with  $m_{3/2}=150\, \rm{GeV}$, $\theta=0.4$,
$\Theta_1=0.9$ and $\tan\beta=2$. In frame $a)$ the values of $B$ and
$\mu$ are assumed to be independent and their magnitudes  are set to
$|B|=100\, \rm{GeV}$ and $|\mu|=600\, \rm{GeV}$. Frame $b)$ shows
the results for the case when electroweak symmetry is assumed to be
broken radiatively.
\label{fig:phi}}
\end{figure}

The dominant sources of CP-violation that are relevant to the electroweak
baryogenesis scenario come from the interactions of the Higgs fields 
with charginos and neutralinos. These interactions couple the Higgsino
and gaugino components of charginos and neutralinos and involve
potentially large CP-violating phases originating from the mixing.
The combined Higgsino source term is then expressed as  
\begin{equation}
{\cal S}_{\tilde{H}} = 3 \gamma_{\tilde{W}} \sin (\varphi_{\mu}) +
\gamma_{\tilde{B}} \sin (\varphi_1 + \varphi_{\mu})
\end{equation}
where 
\begin{eqnarray}
\gamma_{\tilde{W}}&=& |\mu||M_2|g_2^2 v^2 (X) \dot{\beta} (X) 
{\cal I}_{\tilde{W}}, \\
\gamma_{\tilde{B}} &=& |\mu||M_1|g_1^2 v^2 (X) \dot{\beta} (X) 
{\cal I}_{\tilde{B}}. 
\end{eqnarray}
The phase space integrals ${\cal I}_{\tilde{W}}$ and ${\cal
I}_{\tilde{B}}$  can be evaluated in terms of the soft breaking 
parameters entering the chargino and neutralino mass matrices \cite{riotto}.
It is important to emphasize that the phases
factorize from the rest of the source term and enter independently of the 
particular details going into the calculation of the Higgsino thermal 
production rate. For a light superpartner spectrum the produced baryon 
density to entropy density ratio  $n_B/s$ can be as big as $10^{-7}$ 
compared to the observed value of $ 4 \times 10^{-11}$. Recently
identified new terms in the Higgsino thermal production rate \cite{src} can
increase the produced asymmetry even more so the upper limit of
$10^{-7}$ should be viewed as a conservative estimate.

\begin{figure}[t]
\centerline{\psfig{figure=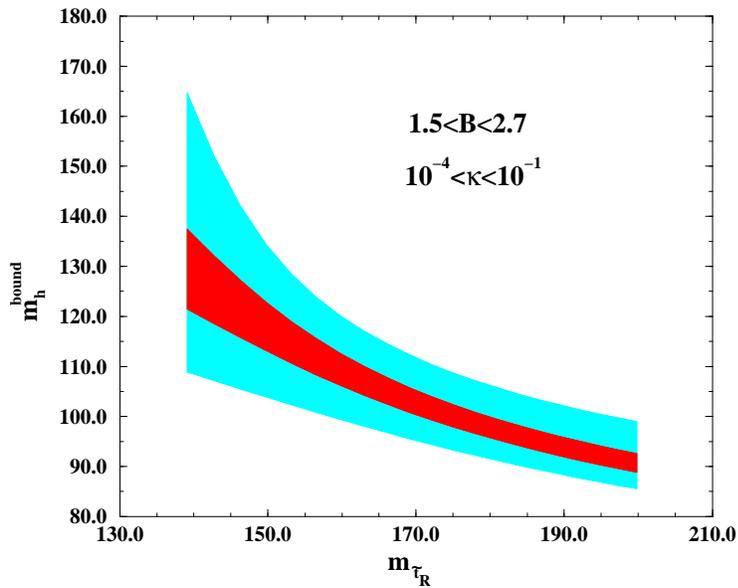,height=3.5in}}
\caption{ 
Dependence of the upper Higgs mass bound on the light right
handed stop mass. The width of the bands results from uncertainties in
the  weak sphaleron parameters B and $\kappa$ which are varied in the
full range. The central band corresponds to B=1.87 while $\kappa$ is varied in
the full range.
\label{fig:mass}}
\end{figure}

Since the baryon asymmetry is overproduced at the time of phase
transition some of it has to be washed out by sphaleron decays in the
broken phase. Defining $\zeta_c \simeq 36 {v(T_c) \over T_c}$
the washout equation \cite{baryo} is for a given value of  $n_B/s$ 
\begin{equation}
\zeta_c - 6 \log \zeta_c - \log \kappa - 9 \log 10 - \log 4.1  + 
\log(\log {n_B/s (T_c) \over 4 \times 10^{-11}}) \gsim 0.
\end{equation} 
The lower limit for $v(T_c)/T_c$ obtained by solving this equation is
related to the soft SUSY breaking parameters determining the position of
the temperature dependent potential minimum. Therefore, it can be
translated into an upper limit on the lightest Higgs mass \cite{mass}
which itself is a function of the right-handed stop mass as shown in
Fig. 2. 
A significant uncertainty is introduced by the estimated range of the weak
sphaleron decay parameters in the broken phase B and $\kappa$.     

The results indicate that by including large CP-violating phases in the 
calculation of the baryon asymmetry the light Higgs masses can be pushed 
towards larger values which can easily accomodate experimental limits
even for relatively heavy ($\sim 200\, \rm GeV$) right-handed stops.
In the case of the Type I string model the EDM cancellation
mechanism requires small values of $\tan\beta$, which constraint the
largest possible light Higgs mass, and Fig. 2 can be used to estimate the 
implied range of the stop mass. Regions with large light Higgs mass
limits and light stops might be relevant in non-minimal supersymmetric
extensions of the Standard Model. 

The idea of the supersymmetric origin of CP violation can be extended
even further. Assuming that the CKM matrix is (approximately) real and
only flavor-independent CP-violating phases occur in the MSSM it is
possible to construct 
a flavor structure for the squark mass matrices at low energies which is 
required for consistency with the observed values of  $\epsilon$,
$\epsilon'/\epsilon$ and the preliminary experimental value of $\sin
2\beta$. At the same time, this framework would lead to theoretical 
predictions such as $\sin 2\beta=-\sin 2\alpha$ which should be
experimentally testable \cite{cp}.

\end{document}